\documentclass[mathleft
]{an}
\usepackage{graphicx}
\usepackage{times}
\begin{document}

\Pagespan{789}{}
\Yearpublication{2006}%
\Yearsubmission{2005}%
\Month{11}%
\Volume{999}%
\Issue{88}%

\title{Study of the effect of metallicity on the amplitudes of Cepheids}

\author{P. Klagyivik\inst{1}\fnmsep\thanks{\email{P.Klagyivik@astro.elte.hu}\newline}
\and  L. Szabados\inst{2}\fnmsep\thanks{\email{szabados@konkoly.hu}\newline}
}
\titlerunning{Effect of metallicity on Cepheid amplitudes}
\authorrunning{P. Klagyivik \& L. Szabados}
\institute{
Department of Astronomy of E\"otv\"os Lor\'and University, 1112 Budapest, P\'azm\'any P. s\'et\'any 1/A, Hungary
\and 
Konkoly Observatory of the Hungarian Academy of Sciences, 1525 Budapest XII, PO Box 67, Hungary}

\received{3 March 2007}
\accepted{15 May 2007}
\publonline{later}


\abstract{%
Results concerning the dependence of photometric and radial velocity amplitudes on metallicity are presented based on about 200 Galactic classical Cepheids pulsating in the fundamental mode. The Galactic distribution of the [Fe/H] value of Cepheids is also studied. We show that the photometric amplitude ratio $A_\mathrm{I}/A_\mathrm{V}$ is independent of metallicity. The observed dependence of this ratio on the pulsation period does not correspond to the theoretical predictions.}

\maketitle

\section{Introduction}

Cepheids are regularly pulsating variables and their radial pulsation occurring with the eigenperiod of the free oscillation gives rise to the existence of various relationships between the period and other physical or phenomenological properties of the star. Widely known among them are the period--luminosity ($P$-$L$) and period--luminosity--colour ($P$-$L$-$C$) relationships which raise Cepheids to a rank of standard candles in establishing the cosmic distance scale.

Though general properties of Cepheids and the pulsation itself can be related to the value of the pulsation period, the atmospheric metallicity also plays a non-negligible role in shaping the oscillation.

The influence of the heavy element abundance, $Z$, on the actual form of the $P$-$L$-$C$ relation has been a frequently discussed topic in the literature on Cepheids. The metallicity dependence of Cepheid luminosities was first mentioned by Gascoigne (1974), while the first thorough investigation was performed by Stothers (1988). These early theoretical studies already pointed to the fact that increasing heavy element abundance results in a higher luminosity. At that time, however, this prediction could not be checked by observational data.

The first empirical determination of the effect of the chemical composition on the $P$-$L$ relationship involving individual calibrating Cepheids was attempted by Fry \& Carney (1997). Reliable data on chemical abundances for a\linebreak large number of Cepheids began to accumulate somewhat later, in early 2000s. The spectroscopic abundance determinations indicate that Galactic Cepheids are not homogeneous from the point of view of their heavy element abundance. Nevertheless, when studying the metallicity dependence of the Cepheid distance scale, the external galaxies are always characterized by an `average' value of metallicity (see e.g. Kennicutt et~al. 1998; Sakai et~al. 2004) in lack of direct spectroscopic information on Cepheid abundances.

In order to study the effects of metallicity on Cepheid pulsation, a new project was initiated in which we started studying how the phenomenological properties of Galactic Cepheids depend on the heavy element abundance, $Z$ (Klagyivik \& Szabados 2006). One of our future goals is to find relationships between metallicity and some Fourier coefficients obtained from the decomposition of photometric and radial velocity phase curves in a similar manner as Jurcsik \& Kov\'acs (1996) did for RR Lyrae type variables.

\section{Data}

First, we formed a data base containing the values of the periods, the amplitudes of the light curves in different photometric bands ($U$, $B$, $V$, $R_\mathrm{C}$, $I_\mathrm{C}$), the amplitudes of radial velocity curves and metallicity of galactic Cepheids.

The periods and the amplitudes were taken from the available databases (Moffett \& Barnes 1985, Szabados\linebreak 1997, and Berdnikov et~al. 2000). If the photometric amplitudes were different in the various sources -- the difference exceeded $0.1$~mag -- we determined a new $A_B$ and $A_V$ amplitude on the basis of the original observational data. The order of priority was our own values, Moffett \& Barnes (1985), Szabados (1997), and Berdnikov~et~al. (2000), respectively. In the catalogue published by Berdnikov~et~al. (2000), there are some stars that have $A_{R_\mathrm{J}}$ and $A_{I_\mathrm{J}}$ amplitude in the Johnson system instead of $A_{R_\mathrm{C}}$ and $A_{I_\mathrm{C}}$ in the Kron-Cousins system. In these cases we calculated the corresponding values by Eq.\,(\ref{R_C}) and Eq.\,(\ref{I_C}), respectively\linebreak which are the fits based on the comparison of the amplitudes in the two photometric systems using the stars observed in both systems (93 stars for $R$ and 91 stars for $I$ band).

\begin{equation}
  \label{R_C}
 A_{R_\mathrm{C}} = 1.175(\pm0.006) \times A_{R_\mathrm{J}}
\end{equation}
\begin{equation}
  \label{I_C}
 A_{I_\mathrm{C}} = 1.202(\pm0.011) \times A_{I_\mathrm{J}}
\end{equation}

\noindent where the amplitudes $A_{R_\mathrm{C}}$, $A_{I_\mathrm{C}}$ are in the Kron-Cousins system and $A_{R_\mathrm{J}}$, $A_{I_\mathrm{J}}$ are in the Johnson system.

Due to the current interest in Cepheid metallicities, there are about 200 Cepheids known in our galaxy with direct high dispersion spectroscopic [Fe/H] values. The metallicities were collected from the following pub\-li\-cations: Fry \& Carney (1997), Groenewegen~et~al. (2004), Andrievsky et~al. (2002a, 2002b, 2002c, 2004, 2005), Luck et~al. (2003), Kovtyukh et~al. (2005a), Romaniello et~al. (2005), and Yong et~al. (2006).

To homogenize the metallicities, first we shifted the data published by different authors to a common solar metallicity, $\log{[n(\mathrm{Fe})]}=7.50$ on a scale where $\log{[n(\mathrm{H})]}=12$ \linebreak (Grevesse et al. 1996). Then we calibrated the data given by Fry \& Carney (1997) and Romaniello~et~al. (2005) to the scale corresponding to data given by Andrievsky and collaborators based on the common stars. The values of Yong~et~al. (2006) were calibrated by Luck~et~al. (2006).

\section{Results}

\begin{figure}
\includegraphics[width=57mm,height=80mm,angle=-90]{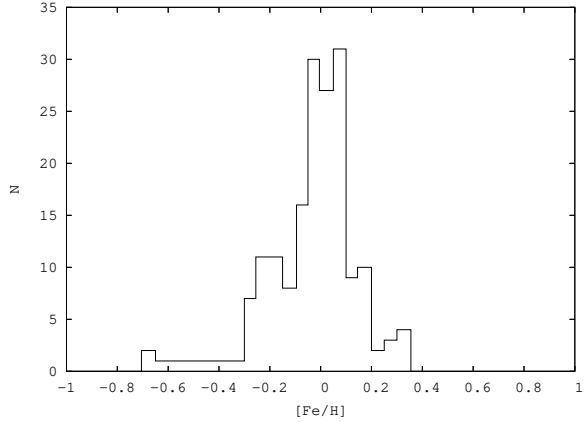}
\caption{The distribution of [Fe/H] values. The width of each interval is $0.05$ dex in the histogram.}
\label{Fe_hist}
\end{figure}

Figure \ref{Fe_hist} shows the distribution of the [Fe/H] ratio. The \linebreak width of each interval is $0.05$ dex in the histogram. The metallicity range is rather narrow, $-0.3$ to $+0.3$ dex and most of the stars have solar like atmospheric chemical abundance. Only a few Cepheids have `extremely' subsolar \linebreak metallicity, below $-0.3$ dex. The error of the individual \linebreak $\mathrm{[Fe/H]}$ values is approximately $\pm0.1$ dex. Therefore the double peak at solar abundance and the valley near $-0.15$ dex might not be real.

\subsection{Effects of binarity}

The most important extrinsic factor that affects the observable photometric pulsation amplitude is the presence of\linebreak companion(s). In the case of radial velocity data we can separate the radial velocity changes caused by the pulsation and the binary orbit, but it is impossible to decompose the two effects in the photometric data. If a Cepheid has a companion it can decrease the photometric amplitude dramatically adding a constant intensity to the maximum and minimum brightness. Moreover the effect of a bright, blue companion is quite different from that of a faint, red one. If we know the luminosity and the temperature of the companion star, we can correct the photometric amplitudes, but mostly this is not the case. Therefore when considering intrinsic photometric amplitudes we have to separate the binary Cepheids from the solitary ones.

\subsection{Distribution of [Fe/H] in the Galaxy}

A number of recent papers investigated the radial metallicity distribution in our Galaxy (Caputo et~al. 2001; Andrievsky et~al. 2002a, 2002b, 2002c, 2004; Luck et~al. 2003, 2006; Kovtyukh et~al. 2005b). All these showed that the metallicity (including iron content) decreases significantly outwards from the center of the Galaxy.

\begin{figure}
\includegraphics[width=57mm,height=80mm,angle=-90]{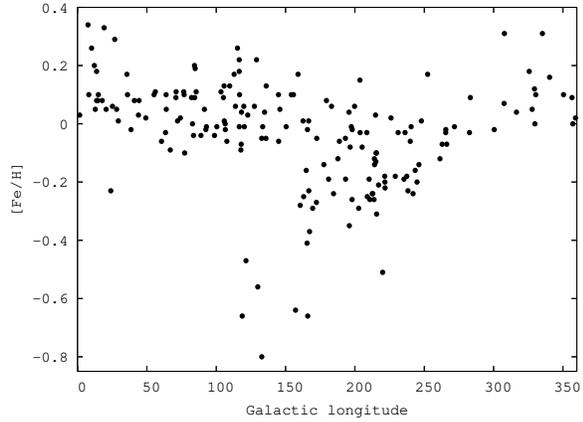}
\caption{The value of [Fe/H] vs. galactic longitude.}
\label{Fe_gal}
\end{figure}

We plotted the [Fe/H] as the function of the galactic longitude (Figure \ref{Fe_gal}). It can be seen, that the iron content is quite uniform ($-0.1 - 0.2$ dex) in all directions with only two exceptions. On the one hand, as expected, the highest values are in the direction of the center of the Galaxy. On the other hand there is a sudden fall between $l=150^{\circ}$ and $l=250^{\circ}$ which includes the anticenter direction. The middle of the decrease is at $l=200^{\circ}$ and not at $l=180^{\circ}$. There are some stars with extremely subsolar atmospheric abundance (below [Fe/H] $= -0.4$). They are located around $l=150^{\circ}$ in the direction of the Perseus spiral arm of the Milky Way.

\subsection{Period and metallicity dependences of the photometric {\boldmath $B$} amplitude}

It is known that the amplitudes of Cepheids depend on the pulsation period in a complicated manner. Namely, a maximum amplitude belongs to every period for all photometric bands and for the radial velocity, as well. Under this upper envelope the distribution of the points is uniform. The envelope was determined in various ways. Efremov (1968) fitted only straight lines to the available limited sample. Later on, Eichendorf \& Reinhardt (1977) developed a method for constructing envelopes to groups of points. By dividing the period range into intervals of equal number of points, they determined a maximum amplitude for all intervals and fitted a polynomial to these data.

We followed almost the same way to determine an upper envelope to our diagram. But we divided the period range into equal width intervals ($0.05$ in $\log{P}$) and a weight was assigned to each interval according to the number of stars involved, and we separated the short and long period Cepheids and fitted third-order polynomials, because a fit to the whole period range was not suitable at the amplitude dip near $\log{P} = 1.0$.

Using these upper envelopes we determined the `amplitude defect', i.e. the magnitude difference between the maximum amplitude corresponding to the given pulsation period and the actual amplitude of each Cepheid. In order to avoid introducing a bias, only solitary Cepheids pulsating in the fundamental mode have been studied. We searched for metallicity effect in the amplitude deviations.

\begin{figure}
\includegraphics[width=57mm,height=80mm,angle=-90]{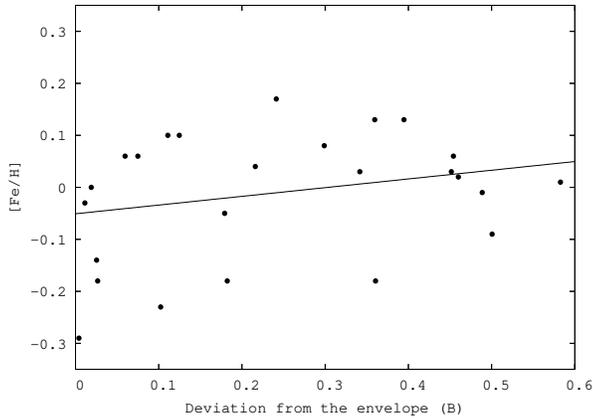}
\caption{$B$ band photometric amplitude defect (see text) vs. [Fe/H] for solitary Cepheids pulsating in the fundamental mode.}
\label{deviation}
\end{figure}

As is seen in Figure~\ref{deviation} there is a tendency that Cepheids with higher iron abundance pulsate with smaller photometric amplitude, in contradiction with the result of a previous study (Paczy\'nski \& Pindor 2000) based on the average metallicity of the LMC and the SMC and some arbitrarily selected Cepheids within them.

\subsection{Ratio of {\boldmath $I$} and {\boldmath $V$} amplitudes}

Tanvir (1997) studied the empirical ratio of the photometric amplitudes in the $I$ (Cousins) and the $V$ (Johnson) bands, and found it to be equal to 0.6. Bono et al. (2000) published the results of theoretical calculations for the \linebreak $A_\mathrm{I}/A_\mathrm{V}$ ratio for fundamental mode pulsators. They studied this ratio for three metallicity values ($Z=0.004$, $Z=0.008$ and $Z=0.02$) and found that it has a mild dependence on metallicity. The mean ratio ranges from $0.64\pm0.03$ at $Z=0.004$ to $0.65\pm0.02$ at $Z=0.02$ and this ratio is constant in average in the whole period range for all three metallicities. The values are somewhat larger than the empirically determined one.

\begin{figure}
\includegraphics[width=57mm,height=80mm,angle=-90]{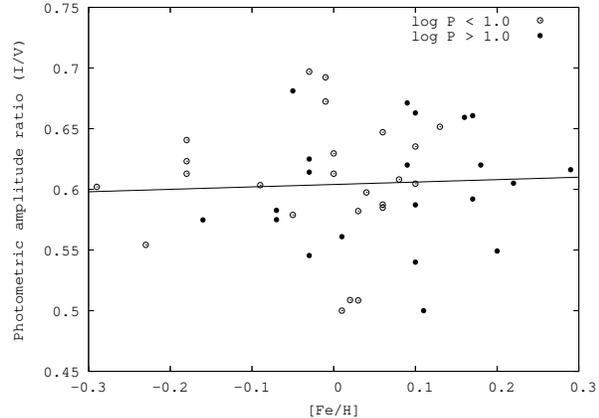}
\caption{Photometric amplitude ratio $A_\mathrm{I}/A_\mathrm{V}$ as the function of [Fe/H]. Open circles shows the short period ($\log{P} < 1.0$) Cepheids, while the filled circles represent the long period ($\log{P} > 1.0$) Cepheids.}
\label{I_V_01}
\end{figure}

\begin{figure}
\includegraphics[width=57mm,height=80mm,angle=-90]{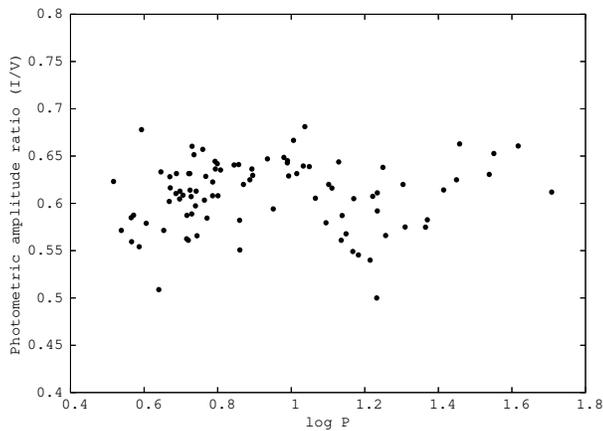}
\caption{Photometric amplitude ratio, $A_\mathrm{I}/A_\mathrm{V}$ vs. pulsation period.}
\label{I_V_02}
\end{figure}

In Figures \ref{I_V_01} and \ref{I_V_02} we present dependence of the $A_\mathrm{I}/A_\mathrm{V}$ ratio on [Fe/H] and period, respectively, based on our new dataset of photometric amplitudes and metallicities. We\linebreak plotted only the solitary and fundamental mode Cepheids because of binarity effects and because the models were calculated for fundamental mode pulsators. The straight-line fit in Fig.~\ref{I_V_01} corresponds to the equation:

\begin{equation}
  \label{I_V_fit}
  A_\mathrm{I} / A_\mathrm{V} =  0.604(\pm 0.008) + 0.02(\pm 0.04) \times \mathrm{[Fe/H]}
\end{equation}

It can be seen, that the error of the slope of the linear fit is twice the slope, so no metallicity dependence can be established. The average is $0.606$, which is slightly lower than the theoretically predicted one and is in good agreement with the previous empirical value from Tanvir (1997).

In Fig.~\ref{I_V_02} we plotted the $A_\mathrm{I}/A_\mathrm{V}$ ratio as the function of the pulsation period. It can be seen, that near $\log{P} = 1.0$ the ratio increases, and after a local minimum at $\log{P} = 1.2 - 1.3$ it shows an increasing tendency again. Bono et~al. (2000) presented model calculations of the period dependence of this ratio (cf. Figure 8 in Bono et al. 2000). The [Fe/H] ratios in our database are solar like ranging from $-0.3$ to $+0.3$ dex with an average of $-0.02$. The agreement between models and observations is good, but there are some differences. The points of the model for $Z=0.02$ are very close to $A_\mathrm{I}/A_\mathrm{V}=0.65$ with small scatter, while the observed values are generally lower, the average is $0.61$ and the model calculations do not show the period dependent pattern seen in Fig.~5. The mean value of the $A_\mathrm{I}/A_\mathrm{V}$ amplitude ratio corresponds to the result predicted by the $Z=0.008$ model, even though it is known that the metallicity of Galactic Cepheids is substantially larger.

\subsection{Other results}

We investigated the metallicity dependence of the slope parameter (a numerical value characterising the wavelength dependence of the $U$, $B$, $V$, $R_\mathrm{C}$ band photometric amplitudes, see Klagyivik \& Szabados, 2006) and the ratio of the radial velocity amplitude and the $B$ band photometric amplitude. We used only stars without known companions, because binarity biases both the slope parameter and the amplitude ratio. While the radial velocity amplitude can be decomposed into pulsational and orbital terms, it is impossible to take into account the photometric effects of the companion star without knowing its brightness and color.

Separating the short and long period fundamental mode Cepheids and the first overtone pulsators, we found that the amplitude ratio ($A_{\mathrm{V_{rad}}}/A_\mathrm{B}$) decreases towards higher [Fe/H] values, and all three groups of stars follow the same relation. In the case of the slope parameter there are some differences between the groups. The short period fundamental mode Cepheids and the first overtone ones behave quite similarly. They show only a mild dependence on metallicity, but the long period pulsators show a more distinct relation and the slope parameter increases with increasing metallicity.

\section{Summary}

Some results concerning the relationships between the spectroscopically determined iron abundance and the pulsational amplitudes (in $U$, $B$, $V$, $R_\mathrm{C}$, $I_\mathrm{C}$ bands, and for radial velocity variations) have been presented. The results indicate that the abundance of the heavy elements in the Cepheid atmosphere has an influence on the observable amplitude of the photometric and radial velocity variations. The higher the iron abundance, the smaller the photometric amplitude. A more detailed analysis of the metallicity dependence of the amplitudes and related parameters will be published in forthcoming papers.

\acknowledgements

Financial support from the OTKA T046207 grant is gratefully acknowledged. We are indebted to the referee, Dr. G\'eza Kov\'acs for his useful comments.


\end{document}